\documentstyle[11pt,paspconf,epsfig]{article}

\begin{document}




{\title { Cosmological Signatures in Temporal and Spectral  
 Characteristics of Gamma-Ray Bursts}
  
\author{Vah\'e Petrosian and Nicole Lloyd}}
\affil{Center for Space Science and Astrophysics,  
 Varian 302c, Stanford University,
 Stanford, CA 94305-4060. \altaffilmark{1}} 
 
\author{Andrew Lee} 
\affil{Stanford Linear Accelerator Center, Stanford University,   
 Stanford, California 94309. \altaffilmark{2}
} 

\altaffiltext{1}{Also Astronomy Program and Department
of Physics.}
\altaffiltext{2}{Work supported by
Department of Energy contract  DE--AC03--76SF00515.}

\begin{abstract}

There have been several suggestions of the existence of cosmological redshift
signatures in the temporal and spectral characteristics of gamma-ray bursts.
However, recent discoveries of afterglows and redshift measurements indicate 
the
presence of broad ``luminosity functions'' which may overwhelm such weaker
cosmological signatures.  The primary goal of this paper is to determine if 
the
intrinsic and cosmological dispersions can be separated.  We have expanded the
search for cosmological signatures to several other temporal and spectral
features, have determined correlations which could arise from the cosmological
redshift of the sources and have carried out tests to determine if the 
observed
correlations can be due to cosmology.  We find that the intrinsic dispersions
are the dominant factors and that detection of cosmological signatures must
await accumulation of a much larger number of identification with galaxies and
measurements of redshifts.

\end{abstract}

\section{INTRODUCTION}

Gamma-ray Bursts (GRBs) show a wide dispersion in their temporal and spectral
characteristics and show a variety of correlations between these.  Now that 
the
association of some GRBs with external galaxies at high redshifts is firmly
established, the question arises whether these dispersions and relations are 
the
consequence of their redshift distribution or are intrinsic to the sources
related to the physics of the emission processes.  The first indication from a
handful of sources with known redshifts is inconclusive in this regard.  
Figure
1 shows the Hubble diagram for GRBs and their afterglows with known redshifts:
On the right we give the variation of the peak (or representative) fluxes, $f_p$ 
(or
$\bar f$), and on the left the fluences $F$ at different photon energies.  
There
is no obvious Hubble relation for any of these measures of burst strength in
most cosmological models.  The small number of GRBs with known redshift $z$
makes it difficult to draw any significant conclusion except that the
``luminosity functions" for all these measures are very broad and mask the 
weaker
cosmological signature.  More redshifts are needed to unravel these two 
effects
from each other.

\begin{figure}
\leavevmode
\centerline{
\psfig{file=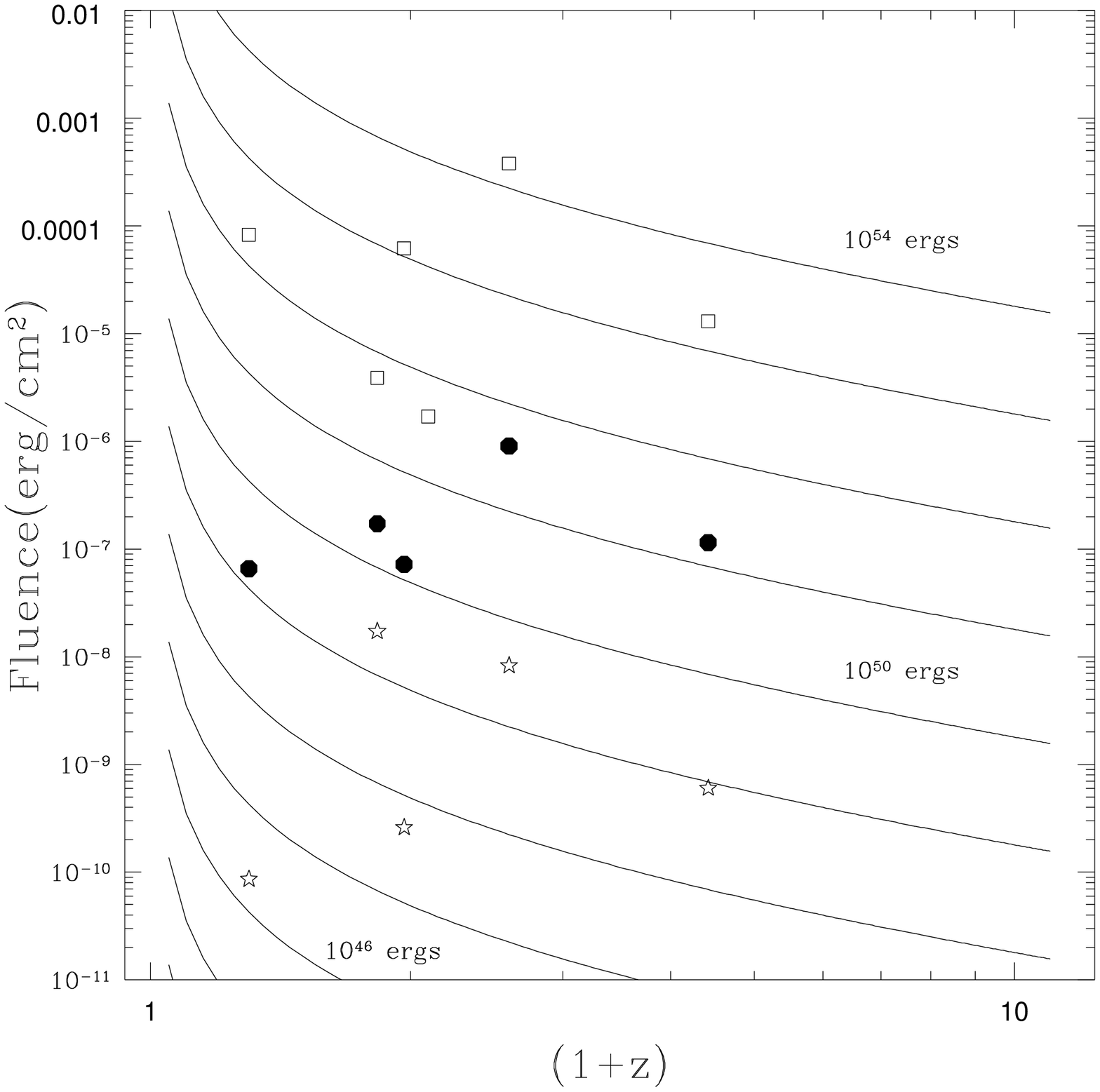,width=0.5\textwidth,height=0.5\textwidth}
\psfig{file=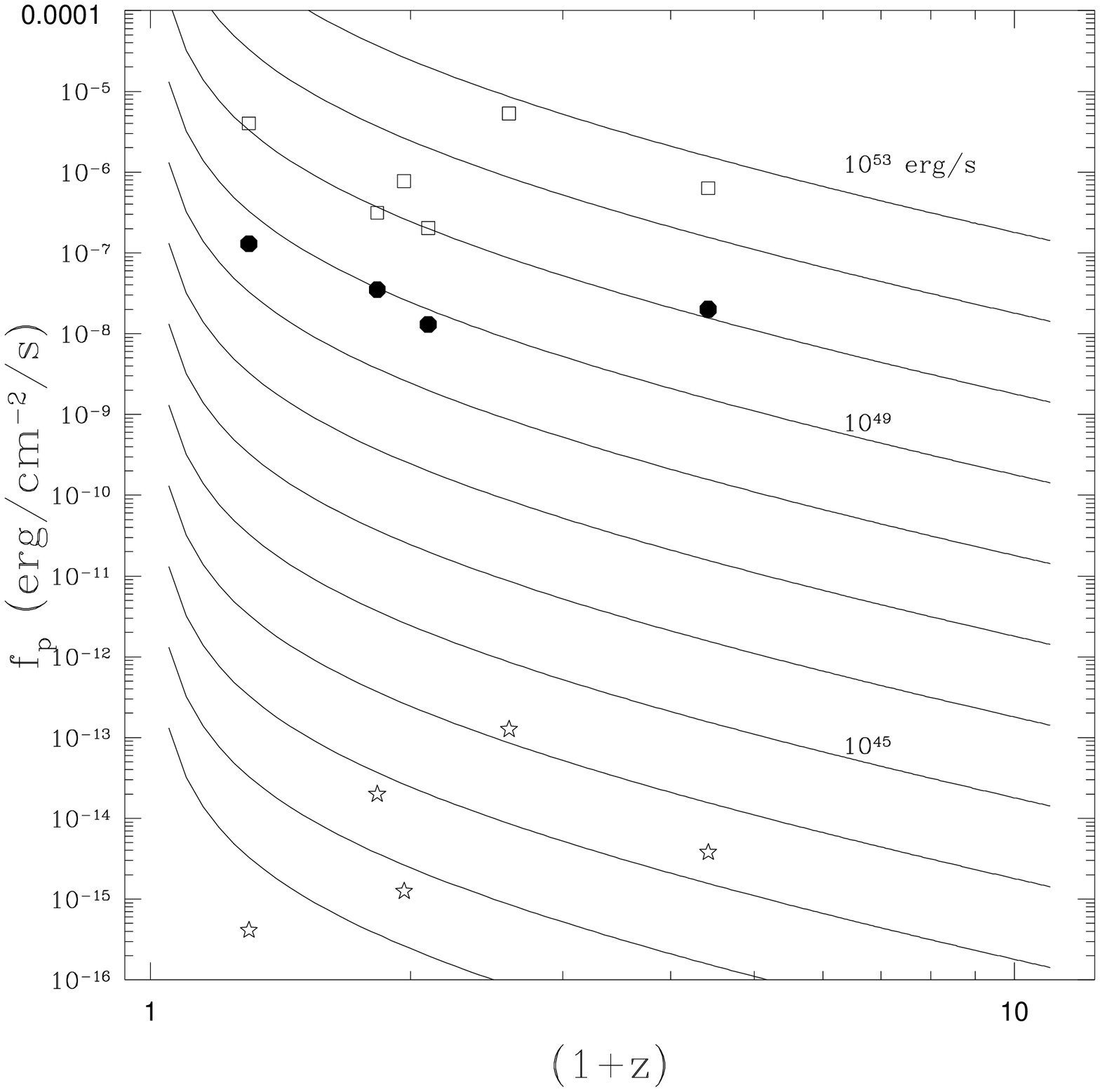,width=0.5\textwidth,height=0.5\textwidth}
}
\caption{\, {\bf  Left Panel:\,} Hubble diagram for fluences at gamma-ray 
(squares), X-ray (filled circles) and optical (stars) ranges for GRBs with known 
redshifts. The solid lines are the expected relation in the 
Einstein-de Sitter cosmological model (with Hubble constant
of 60 km/(s Mpc)) for indicated total 
radiated energy.
{\bf Right Panel:\, }Same as the left panel but for peak or some representative 
fluxes. The X-ray fluxes are the values 8 hours after the burst, and
the optical fluxes are either the peak or the earliest detection
flux. The curves are 
labeled by the value of the luminosity.
}
\label{hubble}
\end{figure}

There have been several attempts to look for cosmological signatures in other 
characteristics of GRBs. Notable among these are the so-called time dilation 
effect as measured by the peak flux-duration correlation (see, e.g., Norris et 
al. 
1994, 1995; J. Bonnell in these proceedings) and 
the 
spectral redshift effect as measured by the correlation between $f_p$ and 
$E_p$, the 
peak or break photon energy of the $\nu F_{\nu}$ spectrum (Mallozzi et al. 
1996, 
1998).
The aim of this paper is  to describe results from further explorations of 
these 
signatures  and to examine whether or not we can shed more light on the 
question 
of intrinsic versus cosmological origins of these relations.

In \S 2 we describe the statistical methods we use for proper accounting of 
the
selection biases and data truncations and how we determine these biases for 
any
physical quantity measured by BATSE.  In \S 3 we describe temporal relations
based on Andrew Lee's thesis and in \S 4 we discuss results from Nicole 
Lloyd's
thesis on spectral relations.  In \S 5 we return to Figure 1 and give a brief
summary and draw some conclusions.

\section{THE STATISTICAL METHODS}

\subsection{Correlations and Distributions}

The statistical problem at hand is to first determine the degree of the {\bf
correlation} between two measured quantities, say $y$ and $z$, and then
determine their univariate {\bf distributions} $\rho (z)$ and $\phi (y)$ from an
observed bivariate distribution $\psi(y,z)$ which suffers from selection 
biases
and is subject to multiple truncations.  The left panel of Figure 2 shows some
generic truncations.  The distribution may be truncated parallel to the axis
(dotted lines) which can be referred to as {\em untruncated} because there is no
bias within the observed ranges.  More interesting cases are when the
truncations are not parallel to the axis.  The data may suffer a one-sided
truncation from below (solid curve) or above (dashed curve), truncated both 
from
above {\bf and} below.
The most general truncation 
is
when each data point, say [$y_i, z_i$], has its individual upper and lower
limits, $y_i^-<y_i<y_i^+$ and $z_i^-<z_i<z_i^+$, as shown by the large cross 
for
one point.  In several papers (Petrosian 1992, Efron \& Petrosian 1992 and 
1999)
we have developed new methods for dealing with all of these situations.  These
are essentially non-parametric methods which avoid the usual arbitrary binning
and the consequent loss of data.  Here we give only a brief description of the
methods.  For further discussion and for examples of applications of these
methods to quasar surveys and GRBs we refer the reader to Maloney \& Petrosian
(1999) and Lloyd \& Petrosian (1999; hereafter {\bf LP99}), respectively.

{\it (a) Correlations:}  Suppose $y$ and $z$ are correlated such that any
characteristic value of $y$, say its average value, varies with $z$ as $g(z)$.
This would mean that we can write $\psi(y,z)=\phi(y/g(z))\rho(z)/g(z)$, where
$\int_0^\infty \phi(x)dx=1$.  The determination of the correlation function
$g(z)$ is based on the rank order $R_i$ of each source among its {\it
comparable} or {\it eligible} set $J_i = \{ j :  x_j > x_i, \ x_j \in (x_i^-,
x_i^+)\}$, where $x$ stands for either variable.  In the absence of any
correlation ($g(z)=$ constant), these ranks will be distributed uniformly so
that their average or expected values would be $E_i = (N_i + 1)/2$ and their
variances $V_i = (N_i^2 - 1)/12$, where $N_i$ is the number of points in 
$J_i$.
One then defines the test statistic $\tau = { {\sum_i ( R_i - E_i
)}/\sqrt{\sum_i V_i} }$.  This statistic is equivalent to Kendall's $\tau$ 
test
and for independent variables its distribution should be a Gaussian with mean 
of
zero and dispersion of unity.  Thus, $y$ and $z$ will be considered 
uncorrelated
or stochastically independent if $|\tau| < 1$, in which case one may assume 
that
the correlation function is constant ($g(z)=1$, say) and proceed with the
determination of the univariate distributions $\phi (y)$ and $\rho(z)$ using 
the
methods mentioned below.  However, if $|\tau| > 1$ then - at the
$1 \sigma$ level - $y$ and $z$ cannot 
be
considered independent and one may assume that the most likely explanation is
the presence of some correlation ($g(z) \neq $ constant).  One can then
determine the function $g(z)$ parametrically as follows.

Given a parametric form for the correlation function $g_k(z)$ one can perform
the transformation $y_{o,i} (k) = y_i / g_k(z_i)$ and proceed with the
determination of the test statistic $\tau(k)$ for the new variables $y_o$ and
$z$ as a function of $k$.  The most likely value of $k$ is that with $\tau(k) 
=
0$ and the range of $k$ for $1 \ \sigma$ confidence level is $\{ k:  
|\tau(k)|
< 1 \}$.  The right panel of Figure 2 shows an example of our results using 
this
procedure.

{\it (b) Distributions:}  Once the function $g$ and its parameter(s) $k$ are
determined, the original bivariate distribution becomes separable when written 
in terms of
$y_o$ and $z$; $\psi(y,z)=\phi(y_o)\rho(z)$.  The last remaining step is the
determination of the univariate distributions $\phi(y_o)$ and $\rho(z)$.  As
shown by Petrosian (1992) all non-parametric methods for this task, in the 
case
of one sided truncation, lead to a generalized form of the Lynden-Bell's (1971)
$C^-$ method.  Our new methods (Efron \& Petrosian 1999) have modified this
procedure so that it is applicable to the general case of the arbitrary
truncation described above. 

\begin{figure}
\leavevmode
\centerline{
\psfig{file=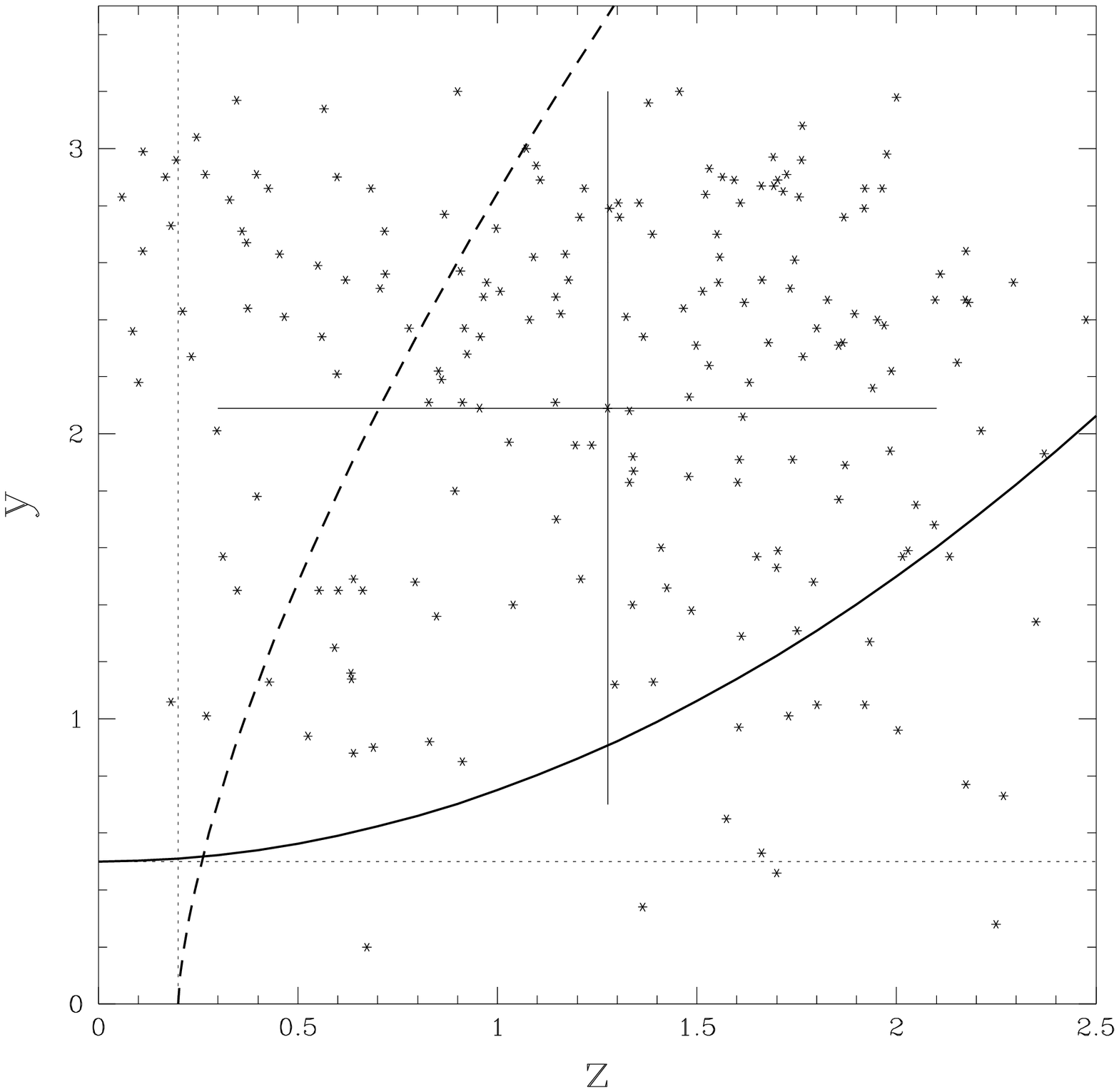,width=0.5\textwidth,height=0.5\textwidth}
\psfig{file=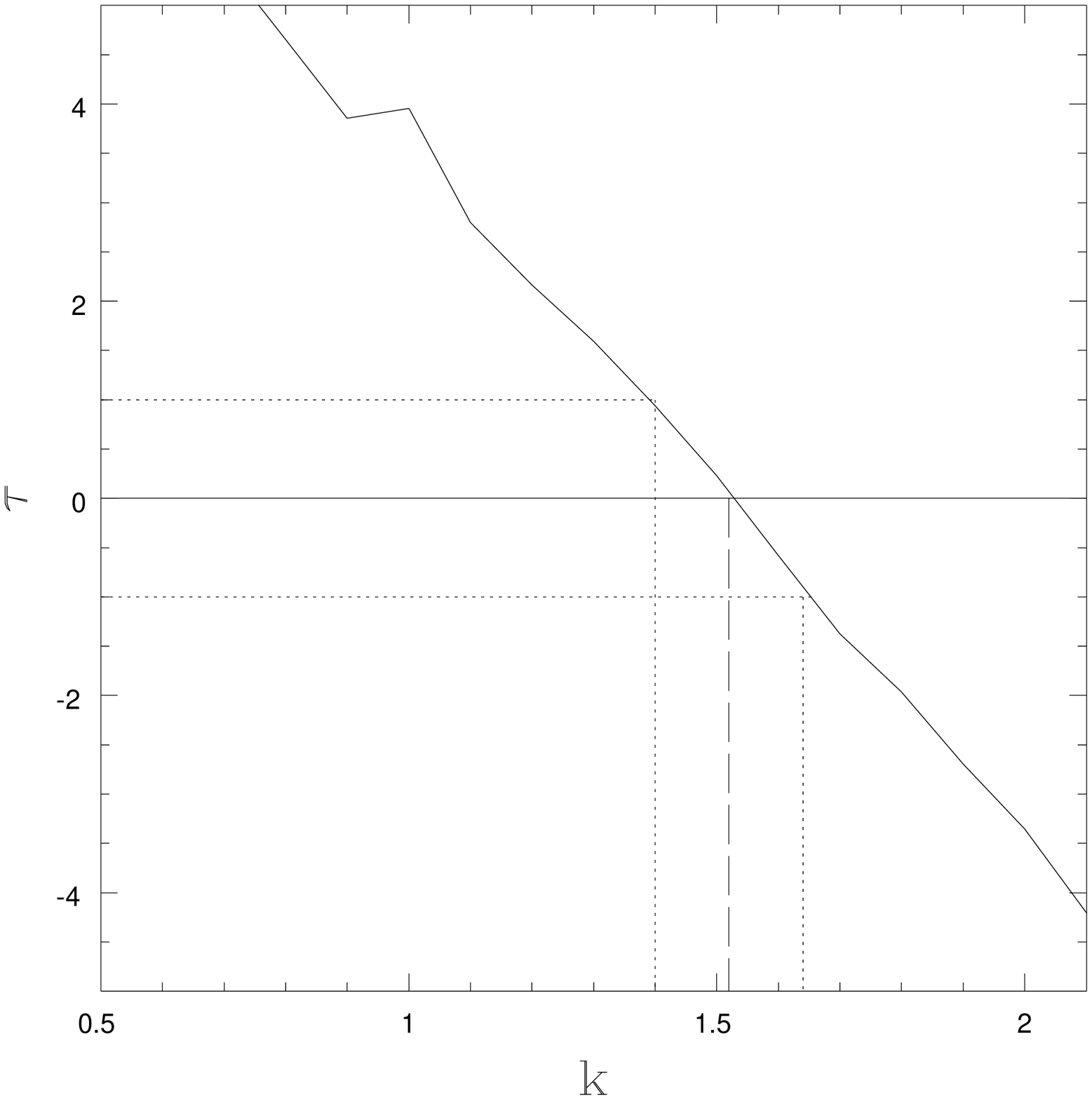,width=0.5\textwidth,height=0.5\textwidth}
}
\caption{\, {\bf  Left Panel:\,} Demonstration of various types of data 
truncations: Parallel to axis (dotted lines), from below (the solid curve), 
from 
above (the dashed curve), and a general truncation when each data point has 
its 
specific observable range (shown by the  cross for only one of the points).
{\bf Right Panel:\, }A determination of the correlation function parameter for 
the 
parametric form $g_{k}(z)=(z) ^ {k}$. The correlation statistic 
$\tau$ is shown as a function of $k$. The solid line at $\tau = 0$ gives
the optimal value $k = 1.52$ and the
dotted lines at $|\tau| = 1$ demonstrate the $ 1 \ \sigma$
range $[1.40,1.64]$.  
}
\label{methods}
\end{figure}

\subsection{Determination of Truncations}

Before we can apply these methods we must accurately account for the
observational selection biases and determine the exact form of the 
truncations.
This task is generally easy for most astronomical samples of sources which are
frequently only flux limited.  However, this is not so simple for transient
sources, in particular for GRBs as observed by BATSE, where many factors such as 
the
trigger criterion, duration, light curve, spectral shape can introduce biases.
In general, instead of a simple ``luminosity function'' one is dealing with a
multivariate distribution of radiant energy, peak luminosity, and spectral and
temporal parameters.  Because of the interrelations between these variables, 
the
selection biases truncate the BATSE data in a complex way.  Fortunately BATSE
has the well defined triggered criterion of peak count (on time scales
$\Delta t=1024, 256$ and 64ms) $C_{max}$ being greater than some minimum value
$C_{min}$.  This can be used to determine the threshold(s) for other physical
quantities measured by BATSE.

One way to carry out this task is by using extensive simulation for each
specific measure (see e.g.  Bloom et al.  1996; Pendleton et al.  1998).  
This
could be very time consuming and computer intensive.  We have developed a much
simpler procedure to account for the {\bf selection biases} and to determine 
the
{\bf thresholds}.  Given the BATSE trigger criterion $C_{max}>C_{min}$, in the
spirit of the $V/V_{max}$ test, we ask:  What is the threshold for a given
observable such that if the burst was as weak as this threshold (e.g., being
farther away) its $C_{max}$ would fall below $C_{min}$?

 It is easy to see that the threshold for any measure of burst strength, say 
the
energy or photon fluences $F$ and $F_{\gamma}$, peak flux $f_p$ or the average
flux ${\bar f}$, is obtained from the following relations:

\begin{equation} 
\label{limits} 
{C_{max}\over C_{min}}\,\, =\,\, {f \over
 f_{lim}}\,\,=\,\,{\bar f\over
\bar f_{lim}}\,\,=\,\, {F_{\gamma}\over F_{\gamma,lim}}\,\,=\,\, {F\over
F_{lim}}.  
\end{equation}

\noindent As shown by Lee \& Petrosian (1996, 1997) the results from this
simple but robust procedure when applied to the fluence $F$ agrees with that of 
the
simulations by Bloom et al.  (1996).  This procedure need not be limited only 
to
measures of burst strength, but can be used for any other measured quantity.
Examples of this are temporal characteristics such as duration or other
parameters describing the light curve, or spectral characteristics such as
$E_p$, and
low and high energy spectral indices $\alpha$ and $\beta$ in any broken power
law spectral form.  For these parameters the thresholds may be more complex;
instead of a single lower limit there may be a lower and an upper threshold in
which cases we use our procedure for two-sided truncated data.  Given a burst 
of
an observed bolometric (in the gamma-ray range) fluence and $E_p$, it is clear
that such a burst would not trigger the BATSE detectors, i.e.
$C_{max}<C_{min}$, if the spectrum was either two soft or two hard.  In other 
words,
$E_{p}$ must be confined in a range  
$E_{p,min}<E_p<E_{p,max}$ in order for that particular burst
to be observed.  For a preliminary application of this procedure 
see
LP99.  In essence {\em the triggering criteria based on counts can be 
translated
to thresholds on any other measured quantity}.  However, care is necessary 
when
dealing with spectral parameters.  The detector response nonlinearities, e.g.
nondiagonal elements in the detector response matrix, DRM, can be important 
and
must be taken into consideration.  We discuss this further below.

\section{TEMPORAL ANALYSES}

Existance of some correlation between peak flux $f_p$ and some measures of 
burst
duration or width is well established (see J.  Bonnell and references cited
there, in these proceedings).  However whether this can be interpreted as a
signature of the cosmological redshift is controversial.  It appears that the
time scale on the rising part of the bursts shows a much smaller or zero
correlation
than the decay time scale (Stern et al. 1997; Mirofanov 1998).  Furthermore, as evident from
Figure 1, the peak flux is not a good measure of distance or redshift.  To 
shed
some light on this controversy we have carried out several similar tests 
looking
for correlations between other measures of burst strength and temporal
characteristics.  These are all outcome of Andrew  Lee's Ph.D.  thesis.  We now 
give
a brief description of the procedures and some of the most relevant results 
from
this thesis.

\subsection{Data and Pulse Fitting Algorithm}

We have used the BATSE Time-to-Spill (TTS) burst data, which
records the times required to accumulate a fixed number of photons in
each of four energy channels. The TTS
data offer variable time resolution, usually finer than any other
BATSE data types except for the time-tagged event (TTE) data, and  
usually can store complete time profiles of bright, long bursts to be
stored in the limited memory on board the CGRO.

We have used the phenomological pulse model of Norris et al. (1996) 
(see also Stern et al. 1997) to 
decompose gamma-ray burst time profiles into
distinct pulses.  In this model, each pulse is described by five
parameters with the functional form

\begin{equation} 
\label{pulse} 
 C(t)\  =\  A \exp{\left(-\left\vert\frac{t
 - t_{\rm max}}{\sigma_{r,d}}\right\vert^{\nu}\right)}\ ,
\end{equation}

\noindent where $t_{\rm max}$ is the time at which the pulse attains its
maximum, $\sigma_{r}$ and $\sigma_{d}$ are the rise and decay times,
respectively, $A$ is the pulse amplitude, and $\nu$ (the~``peakedness''
parameter) gives the sharpness or smoothness of the pulse at its peak.  We 
have
developed an interactive pulse-fitting program that can automatically find
initial background level and pulse parameters using a Haar wavelet denoised 
time
profile, and allows the user to add or delete pulses graphically.  The program
then finds the pulse parameters by using a maximum-likelihood fit for the
gamma-distribution that the TTS spill times follow (for details see  Lee et al. 
1998).

\subsection{Correlations from Burst to Burst}

We first describe correlation between burst strengths and timescales for 
different bursts. This has direct bearing on the time dilation hypothesis.

We use the amplitude $A$ (see eq.[2]) of the highest amplitude pulse 
in each burst as a
measure of the peak flux, and the total counts ${\cal C}=\int 
C(t)dt=\frac{A}{\nu}(\sigma_r+\sigma_d)\Gamma (1/\nu)$ of the pulses as a
measure of the photon fluence of each pulse and the sum of these for the 
fluence of each burst.  The timescales that we
use are the  FWHM duration ($T_{.5}=(\sigma_r+\sigma_d)({\rm ln}2)^{1/\nu}$) of 
the 
highest amplitude pulse in each
burst, and the interval between the peak times of the two highest
amplitude pulses in multiple-pulse bursts $(\Delta T_{1,2})$. The following 
are some of the results relevant to this paper.

(a) We find an inverse correlation between the amplitudes and
the widths of the  highest amplitude pulse of the bursts; Figure 3, left panel. 
This is 
similar to to the ``time dilation'' trend observed by Norris et al. (1994, 1998) 
but the slope of the trend doesn't appear to agree
with the expected effects of cosmological time dilation alone; the 
variation in pulse width is  greater than expected.

(b) We also find an inverse correlation between the highest pulse amplitude
and the time interval between the peaks of the two highest pulses in
each burst; Figure 3, right panel. This weaker variation may be consistent 
with the expected
results of cosmological time dilation.  It is likely that
this correlation is less affected by intrinsic properties of GRBs
or by selection effects than the correlation in item (a) above.  This
agrees with Norris et al. (1996) and Deng \& Schaefer (1998).

\begin{figure}
\leavevmode
\centerline{
\psfig{file=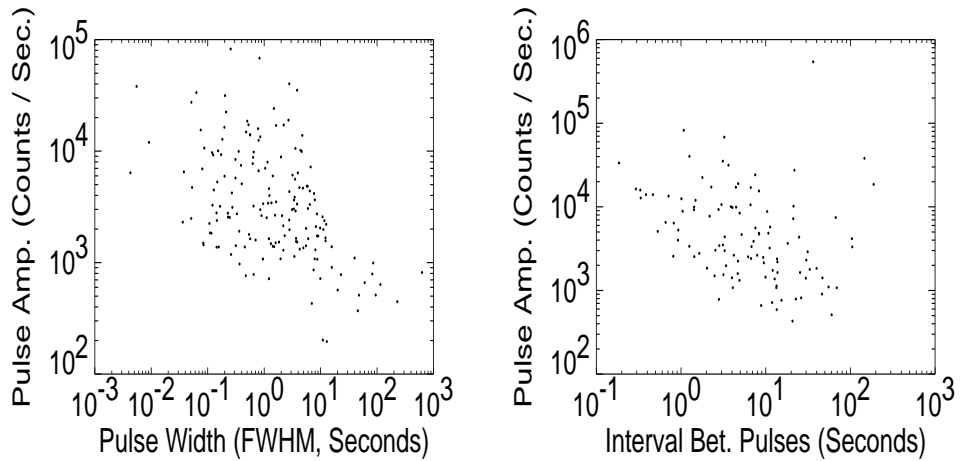,width=1.\textwidth,height=0.5\textwidth}}
\caption{{\bf Left Panel:} Highest pulse amplitude versus pulse width 
($T_{.5}$) of highest amplitude pulse. {\bf Right Panel:} Highest pulse 
amplitude versus interval between two highest pulses $(\Delta T_{1,2})$. 
Energy channel 2, 60-110 keV. Note the clear evidence for anticorrelations.}
\label{ampvst}
\end{figure}

(c) On the other hand, we find a positive correlation between total count
fluences of bursts and the widths $T_{.5}$ of the highest amplitude pulse in
each burst; Figure 4, left panel.  This is in agreement with the fluence-total
duration correlation described by Lee \& Petrosian (1996, 1997).  Clearly in
both cases the cosmological effect (anticorrelation), if any, has been
overwhelmed by the intrinsic correlations.  (But we find no correlation
between total burst count fluence and the interval between the two highest
pulses in each burst; Figure 4, right panel.)

The simulations discussed in section 3.4 show that the observed
correlations between the fluence and amplitude of the highest amplitude pulse
and the two timescales in each burst do not appear to be strongly affected by
the pulse-fitting procedure.

\subsection{Correlations Among Pulses Within Bursts}

For bursts with multiple pulses we can carry out the above tests among pulses
within the bursts.  Any correlation here must be intrinsic to the physical
process of emission and not affected by the cosmological redshifts. We find 
trends qualitatively similiar to those in items (a) to (c) above. The following 
are some of the results relevant to this question.

\begin{figure}
\leavevmode
\centerline{
\psfig{file=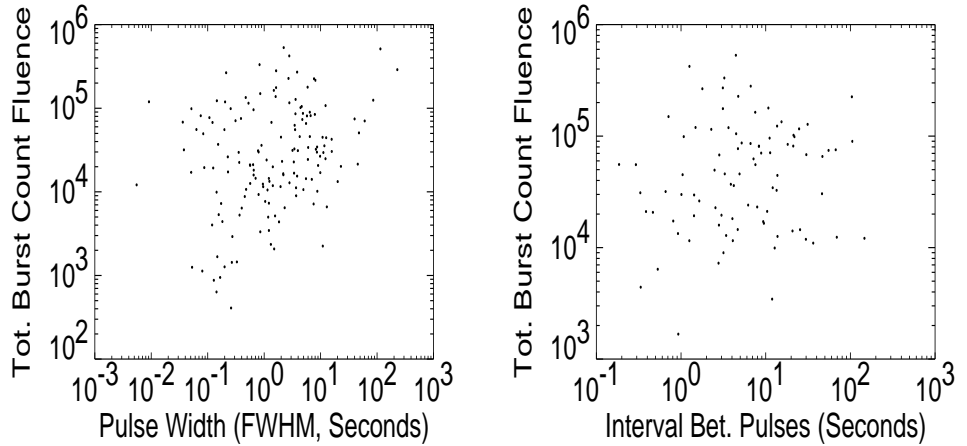,width=1.\textwidth,height=0.5\textwidth}}
\caption{{\bf Left Panel:} Total count fluence ${\cal C}$ of bursts versus pulse 
width 
$T_{.5}$ of highest amplitude pulse. {\bf Right Panel:} Total count fluence 
${\cal C}$ versus interval between two highest pulses $\Delta T_{1,2}$.  
Energy channel 2, 60-110 keV. Note presence of strong correlation in the left 
panel.}
\label{areavst}
\end{figure}

(a) We find that within individual bursts, higher amplitude pulses have a 
strong
tendency to be narrower; the pulse amplitude-duration scatter diagrams tend to 
have more negative than positive slopes. This effect is even stronger for 
bursts with the stronger correlations probabilities between pulse amplitude 
and duration. The scatter diagrams with the individual linear least-squre fit 
to the logarithms of these quantities are shown in the left panel of Figure 5. 
Only one out of about thirty bursts shows a positive slope. The rest show the 
anticorrelation similar to the time-dilation seen among the bursts. Obviously 
the present anticorrelations  cannot be due to cosmological effects and  must 
result from intrinsic properties
of the GRBs themselves, or from selection effects in the
pulse-fitting procedure (see below).

(b) We also find that majority of bursts show a positive correlation between 
the count fluence  and duration of their pulses. The right panel of Figure 5 
shows the trends for bursts with significant correlation probabilities. This 
also is similar to the behavior among bursts which is opposite 
of what is expected from the reshift effect.

\subsection{Biases and Simulation Results}

There are a number of ways in which the pulse-fitting procedure may
introduce biases into correlations between pulse
characteristics.  One is that the errors in the different fitted pulse
parameters may be correlated.  Another is that the pulse-fitting
procedure may miss some pulses by not identifying them above the
background noise.  Still another cause of bias is that
overlapping pulses may be identified as a single broader pulse.

We have tested for these selection effects by generating simulated
burst time profiles using the pulse model in equation (2) with 
randomly generated parameters with distributions that are similar to those 
observed. We add appropriate noise to this data and then fit the simulated 
bursts exactly the same way as the actual data. The following are results from 
comparing the simulated and fitted pulse characteristics.

\begin{figure}
\leavevmode
\centerline{
\psfig{file=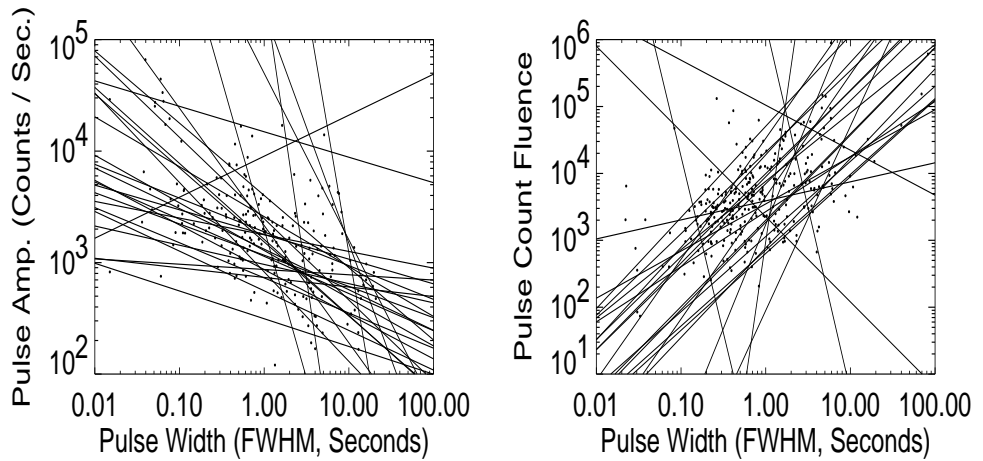,width=1.\textwidth,height=0.5\textwidth}}
\caption{ Pulse amplitudes versus pulse widths ({\bf left panel}) and pulse 
count fluences versus pulse widths ({\bf right panel}) within bursts for 
bursts with strongest correlations. The lines show the fitted power laws to 
pulses in individual bursts with high positive or negative
correlation coefficients. Data from 
energy channel 2, 60-110 keV.}
\label{afwhm}
\end{figure}

For simulated bursts consisting of a single pulse
in both the original simulation and in the fit, the identification of
pulses between the simulation and the fit is unambiguous and
unaffected by the effects of missing pulses. We find that when the fitted
amplitude is larger than the original amplitude, the fitted width
tends to be smaller than the original width, and vice versa (i.e. fluence 
roughly invariant).  This could 
introduce a small bias in the sense of the observed anticorrelation between 
amplitude and duration. This tendency also indicates that there will be no 
bias introduced in the correlations between count fluences and  pulse widths.

Figure 6
shows the amplitude-pulse width relations for the 
simulated bursts before (left panel) and after (right panel) our fitting 
procedure. 
As expected the left panel shows an almost random distributions of the fitted 
slopes while the right panel shows a greater tendency for negative than 
positive slopes (16 vs. 10). This means that the fitting procedure can
create a weak inverse
correlation between pulse amplitude and pulse width within bursts. We do not 
believe that this is sufficiently strong to explain the behavior observed in 
Figure 5 where only one out of thirty well correlated bursts shows positive
slopes.

\begin{figure}
\leavevmode
\centerline{
\psfig{file=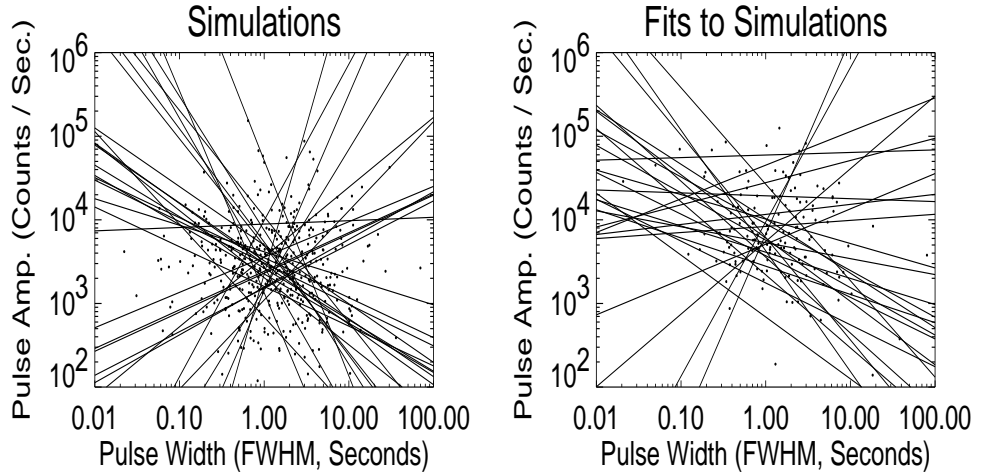,width=1.\textwidth,height=0.5\textwidth}}
\caption{Same as the left panel of Figure 5 except for the raw (left) and fitted (right) simulated 
data.}
\label{sim_afwhm}
\end{figure}

We also find that the fitting procedure slightly weakens the positive 
correlation between the pulse fluence and pulse width so that the positive 
correlation between these quantities found for the actual bursts (right panel 
Fig. 5) is a slight underestimation of the true correlation.

\section{SPECTRAL ANALYSES}

A similar cosmological signature has been claimed to be present in the
correlation between the peak flux $f_p$ and the spectral hardness of GRBs, as
measured by $E_p$ (Mallozzi et al.  1996, 1998; Mitrofanov et al. 1999).  The
question arises again whether this is caused by the redshift of the bursts or
is intrinsic. This is because, as shown in Figure 1, it is not clear if the peak 
flux
provides a good measure of source redshift.  In order to clarify this, we have
looked for correlations between $E_{p}$ and other measures of burst strength,
in particular its total (bolometric) energy fluence, which (according to
models with well defined total released energy, such as mergers or hypernovae)
may be a better redshift indicator.  We use the procedures described in \S 2
to determine the thresholds on these quantities and evaluate the degrees of
their correlations.  We then discuss whether the observed relations could be
 due solely to redshift effects.

\subsection{Bias and Correlations} 

We have previously (LP99) used the burst spectral parameters (for a Band
spectrum) obtained from four channel data.  Here we use spectral parameters
kindly provided by Dr. Robert Mallozzi, who used the software
WINGSPAN to fit a Band spectrum to 16 channel CONT
(continuous) data for a large sample of bursts.  The above mentioned
correlations were obtained from the latter data.  The LP99 sample contains a
complete sample of bursts with known $C_{max}$ and $C_{min}$ values and 
therefore
has a well defined truncation. Unfortunately the Mallozzi sample, 
although
more reliable in the values of the spectral parameters, does not have a well
defined selection criterion.  Because the brightest bursts give the best fits,
this sample includes bright bursts with no known values of $C_{min}$; 
furthermore,
most bursts in this sample 
with a known $C_{min}$ value 
have $C_{max}/C_{min} \gg 1$.  We are 
currently
collaborating with Dr. Mallozzi to determine the biases in this sample in order to
evaluate the significance of the correlations between $f_p$ and $E_p$ that he
and his collaborators have reported.  Without exact knowledge of the selection
bias, we cannot account for data truncations properly or determine 
correlations
accurately.  To circumvent this situation, we chose a subsample with a better
defined selection criterion as follows.  We truncate the available data,
parallel to the axes in the $f_{p}-f_{p,lim}$ or $F-F_{lim}$ plane, so that 
the
above mentioned uncertainty is minimized.  For the observed fluence in the 
range
50-300 keV we select a subsample with $F_{obs}\ge 10^{-6}$ ergs/cm$^{2}$, and
for the total fluence (summed over all four LAD channels; ~20keV-1.5MeV) we
select sources with $F_{sum} \ge 5 \times 10^{-6}$ ergs/cm$^{2}$.  (Note that
$F_{sum}$ is approximately equal to the total fluence of the burst,
$F_{tot}=\int_{E_{min}}^{\infty} F(E)dE$.)  For peak flux, the cut was made at
$f_{p} \ge 3.0$ ph/(cm$^{2}$ s).  Hence, the results presented below are valid
for a narrower range of the parameters near the bright end of the burst
intensity distribution.

As discussed in \S 2 above, we must account for any truncation in the 
variables
we are correlating.  We use the method described in LP99 to get an estimate of
the truncations on $E_{p}$.  In LP99 we pointed out that because BATSE 
triggers
over a finite energy range, if the $E_{p}$  is too far above or too far below 
the trigger range, the burst will 
not
be detected.  Hence, using the burst trigger condition, we can place both an
upper and a lower limit on $E_{p}$.  However, it should be pointed out that this 
does not include
effects of the DRM, which may play an important role in the significance of 
the
truncation.  This is less important for the present sample which has a narrow
distribution of $E_p$'s and consequently requires a small correction due to
instrumental biases (a delta function distribution requires zero correction).
The truncations on the fluence and peak flux are straightforward, since, as
described above, we have made a cut at some well determined value parallel in
the axes in the $F-F_{lim}$ and $f-f_{lim}$ planes.  

We have carried out the correlation test in the samples defined above.  
The results are shown in Table 1.  The last
quantity, $f_{p,trig}$ is the peak photon flux between 50-300keV on the
timescale in which the detector triggered (either 64ms, 256ms, or 1024ms).  
The
fluences are correlated with the burst {\em average} $E_{p}$, while the peak
photon flux is correlated with the value of $E_{p}$ at the time of the peak.
The results are given in terms of the signifcance of the correlation using the
Kendall's $\tau$ test mentioned in \S 2.  The raw result shows the correlation
without accounting for truncation in the variables, while the corrected result
uses the techniques described in \S 2 to account for these truncations.

\begin{center}
\centerline{Table 1}
\centerline{Kendall's $\tau$ Values for Correlations Between Various Strengths  
and 
$E_p$}
\begin{tabular} {lcccr} \hline \hline
   Correlation           & Raw result   & Corrected result & Number of bursts 
\\ \hline
$F_{obs}$   & (+)  5.6 $\sigma$ & (+) 5.5$\sigma$ & 147  
\\ 
$F_{sum}$    & (+)  6.5$\sigma$   &  
 (+)5.8$\sigma$ & 160 
\\ 
$f_{p.trig}$   & (+) 2.5$\sigma$ &  
(+)2.6$\sigma$ & 101
\\ \hline \hline
\end{tabular}
\end{center}

As mentioned above, for our samples (obtained with more strict and rigorous 
selection
criteria), the truncation effects are expected to be small.  This is reflected 
in
the relative values of the raw and corrected $\tau$ values.  We find that 
there
is a strong correlation between the observed fluence and $E_{p}$ as well as  the total
fluence and $E_{p}$.  Similar, but somewhat weaker, correlations are evident in
other samples as well.  However, we find only a moderate correlation between
the peak photon flux and $E_{p}$ for the bright end of the intensity
distribution of the bursts.  This is not the case for the whole sample in
Mallozzi's list for which we find a correlation similar to the results 
reported
previously (Mallozzi et al.  1996, 1998).  However, the most significant
correlation reported by Mallozzi and collaborators comes from bursts below our
cutoff of 3 photons/(cm$^{-2}$ s) where the selection criterion is
uncertain.  These aspects of the problem will be addressed in future
publications.

\subsection{Cosmological Signature?}

The above correlations between  fluences and $E_p$ follow the same trend as 
that expected 
from cosmological effects.
We would like to test if these correlations can be attributed fully to such 
effects.  We will focus particularly on the $F_{sum} \approx
F_{tot}$ results, because the total fluence can be related to the total
radiated energy 
and the redshift of the burst, without any need for the
so-called K-correction; $F_{tot}=E_{rad}/(\Omega_{b} [d_E(\Omega_i,z)]^{2})$, 
where $E_{rad}$ and $\Omega_b$ are the total radiant 
energy  (in the gamma-ray range) and the average beaming sterradians, $\Omega_i$ 
denote the cosmological 
model parameters, and $d_E(\Omega_i,z) =
d_L(\Omega_i,z)/\sqrt (1+z)$ with $d_L$ as the usual bolometric luminosity 
distance. For this task we need to specify  a cosmological model and the 
distribution function of redshift and the intrinsic parameters,
$\Psi(E_{p},\Omega_b,E_{rad},z)$. The beaming angle enters always in 
conjunction with $E_{rad}$. To simplify the matters, in what follows we assume 
a delta function distribution of $\Omega_b$ so that it can be eliminated from 
the distribution function. This amounts to replacing $E_{rad}$ with  
$E_{rad}/\Omega_b$. We also assume no evolution for the intrinsic parameters 
and no intrinsic correlation between $E_{rad}$ and $E_p$. These assumptions 
mean that the multivariate distribution function becomes separable as 
$\Psi(E_{p},\Omega_b,E_{rad},z) = \phi(E_{rad})\zeta(E_p)\rho(z)$. In this 
case the joint distribution of observed $E_p$ and $F_{tot}$ is given by

\begin{equation}\label{diffdist}
	{d^2N(E_{p}, F_{tot}) \over dE_pdF_{tot}} =  \int_{0}^{\infty} dz 
(dV/dz) \rho (z) [d_E(\Omega_i,z)]^{2} \phi(F_{tot}[d_E(\Omega_i,z)]^{2})
(1+z)\zeta(E_{p}(1+z)),
\end{equation}

\noindent where $dV/dz$ is the differential of the comoving volume up to $z$.
From this we can compute the individual distributions, the
average value of $E_{p}$ as a function of $F_{tot}$ or vice versa.  For 
example,

\begin{equation}\label{avEp}
	{\bar E}_p(F_{tot}) = \frac{\int dE_p E_p [d^2N(E_{p}, 			
         F_{tot})/dE_pdF_{tot}]}{dN(F_{tot})/dF_{tot}},
\end{equation} 
where

\begin{equation}\label{Fdist}
	dN(F_{tot})/dF_{tot} =\int dE_p [d^2N(E_{p},F_{tot})/dE_pdF_{tot}].
\end{equation}

We try various plausible models for the functions $\zeta$, $\phi$, 
and $\rho$.
We then compute the expected ${\bar E}_p(F_{tot})$,
remove this cosmological correlation from the data by the transformation
$E_{p}^{'} = E_{p}/{\bar E}_p(F_{tot})$,  and see if we
are left with any correlation between the observed $F_{tot}$ and $E_{p}^{'}$ 
distributions. 
Only if none remains, then  can we attribute the correlation between $F_{tot}$ 
and $E_{p}$
to cosmological effects alone.
We assume that $\phi$ obeys either a delta function (standard candles) or a 
power law with spectral index $\beta$ in the radiated
energy, $\zeta$ is a Gaussian in the intrinsic (i.e. rest frame)
value of $E_{p}$  with a mean 
of $Q$ and dispersion $\sigma_E$,  and $\rho$ is either a constant or follows
the star formation rate. We present results for the Einstein de-Sitter 
cosmological model which are qulitatively similar to that of several other 
models that we have explored. 

\begin{figure} 
\leavevmode
\centerline{
\psfig{file=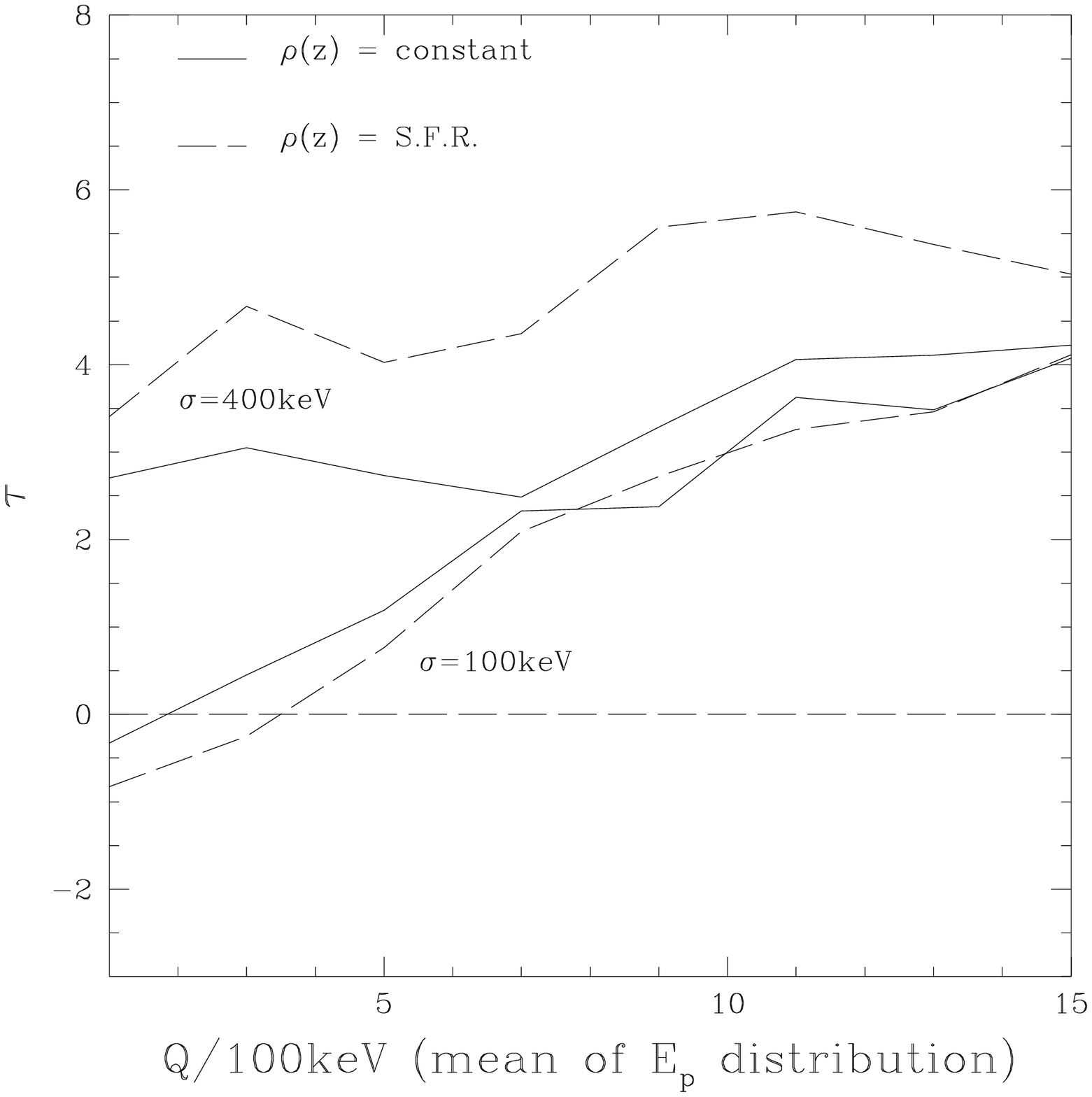,width=0.5\textwidth,height=0.5\textwidth}
\psfig{file=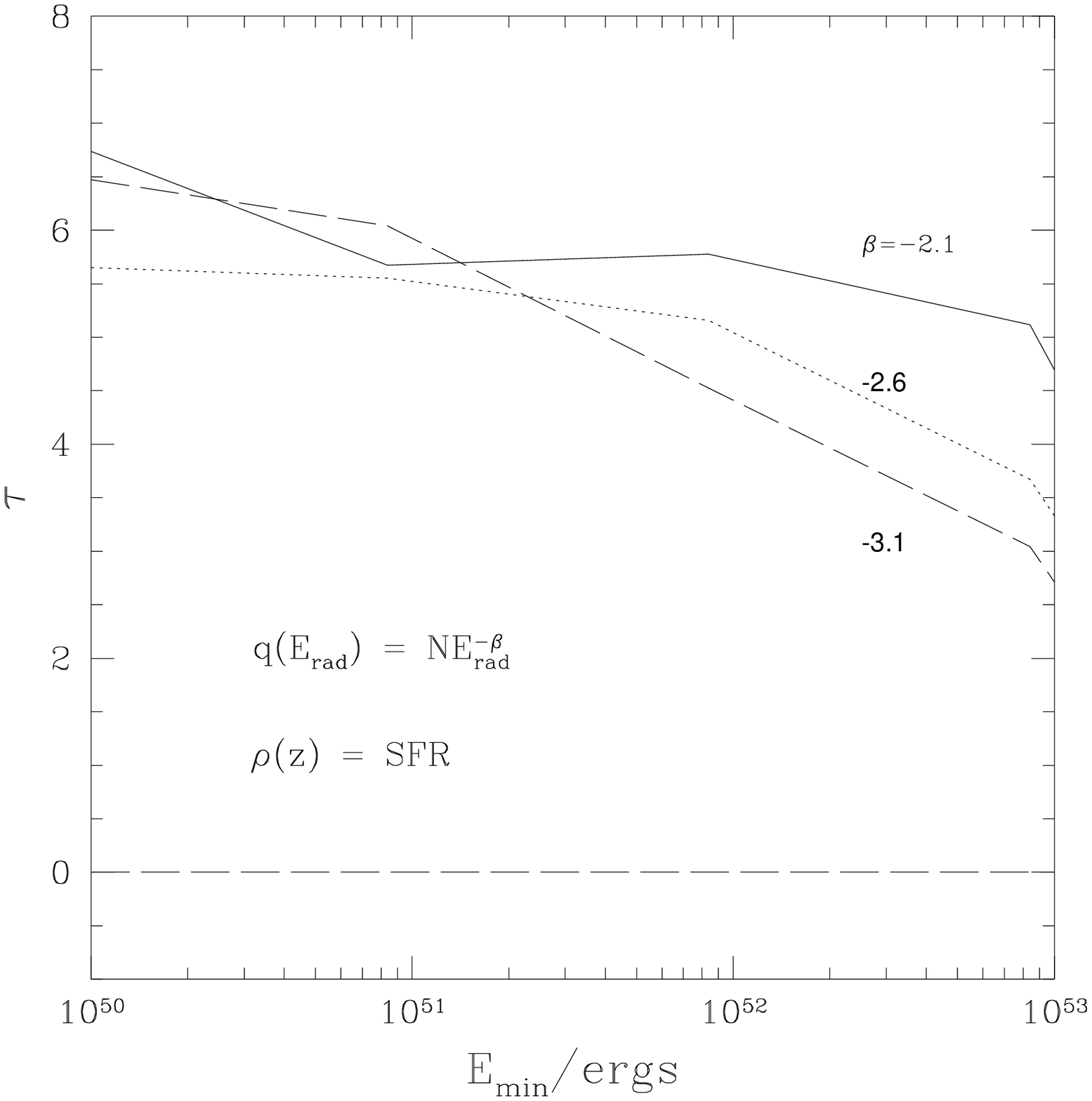,width=0.5\textwidth,height=0.5\textwidth}
}
\caption{\, {\bf  Left Panel:\,} The $\tau$ statistic as a function  of $Q$, the 
mean intrinsic value of $E_p$ (assumed to have a Gaussian distribution), for 
two different values of the dispersion $\sigma_E$ and rate evolution 
$\rho(z)$. The distrubution of $E_{rad}$ is assumed to be a delta function at 
$10^{53}$ ergs.
Note that for a constant rate the observed correlation between $F_{tot}$ and 
$E_p$ can be due to cosmology if the mean and the dispersion values of $E_p$ are 
low.
{\bf Right Panel:\, } Same as the Left panel except for a power law 
distribution of $E_{rad}$ with spectral index $\beta$ and as function of the 
minimum value of this distribution. Here we assume $Q=600$ keV, $\sigma_E=500$ 
keV, $\rho$ proportional to the star formation rate  and a Hubble constant of 
60 km/(s Mpc).
}
\label{tautests}
\end{figure}

Our general conclusion is that, for plausible values of the parameters for 
this
set of distributions, the correlation cannot be attributed to cosmological
effects alone.  This is illustrated in the two panels of Figure 7.  The
cosmological effects can account for all of the observed correlation for model
parameters for which the statistic $\tau=0$.  As evident
from these curves, for most of the plausible combinations of distribution
parameters, $\tau$ remains well above one so that the correlation seen in the 
data cannot be accounted for by
cosmological effects alone.  After the removal of the cosmological
contribution to the correlation, we are still left with a positive correlation
except for a delta function distribution of the radiant energy and a narrow 
intrinsic
distribution of $E_p$'s with a low mean value.  As seen from the left panel of
Figure 1 the first of these cannot be true and the other requirements are also
unreasonable.  One explanation of this is that there is an intrinsic correlation between
total fluence and $E_{p}$.  The degree of this correlation, of course, depends
on how we model the distribution of the burst parameters.

\subsection{Distributions of $E_p$ and $F_{tot}$}
 
\begin{figure} 
\leavevmode
\centerline{
\psfig{file=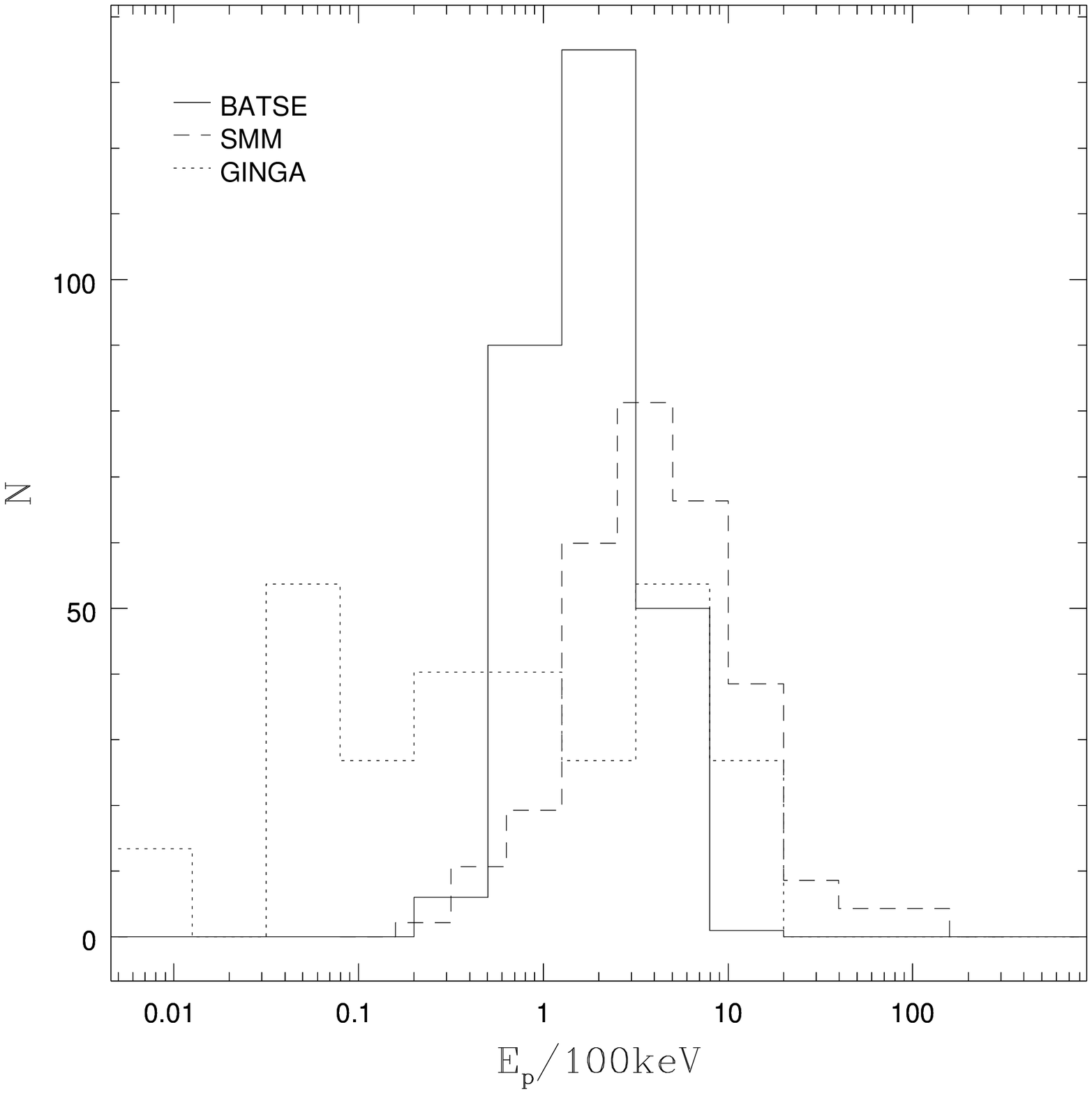,width=0.5\textwidth,height=0.5\textwidth}
\psfig{file=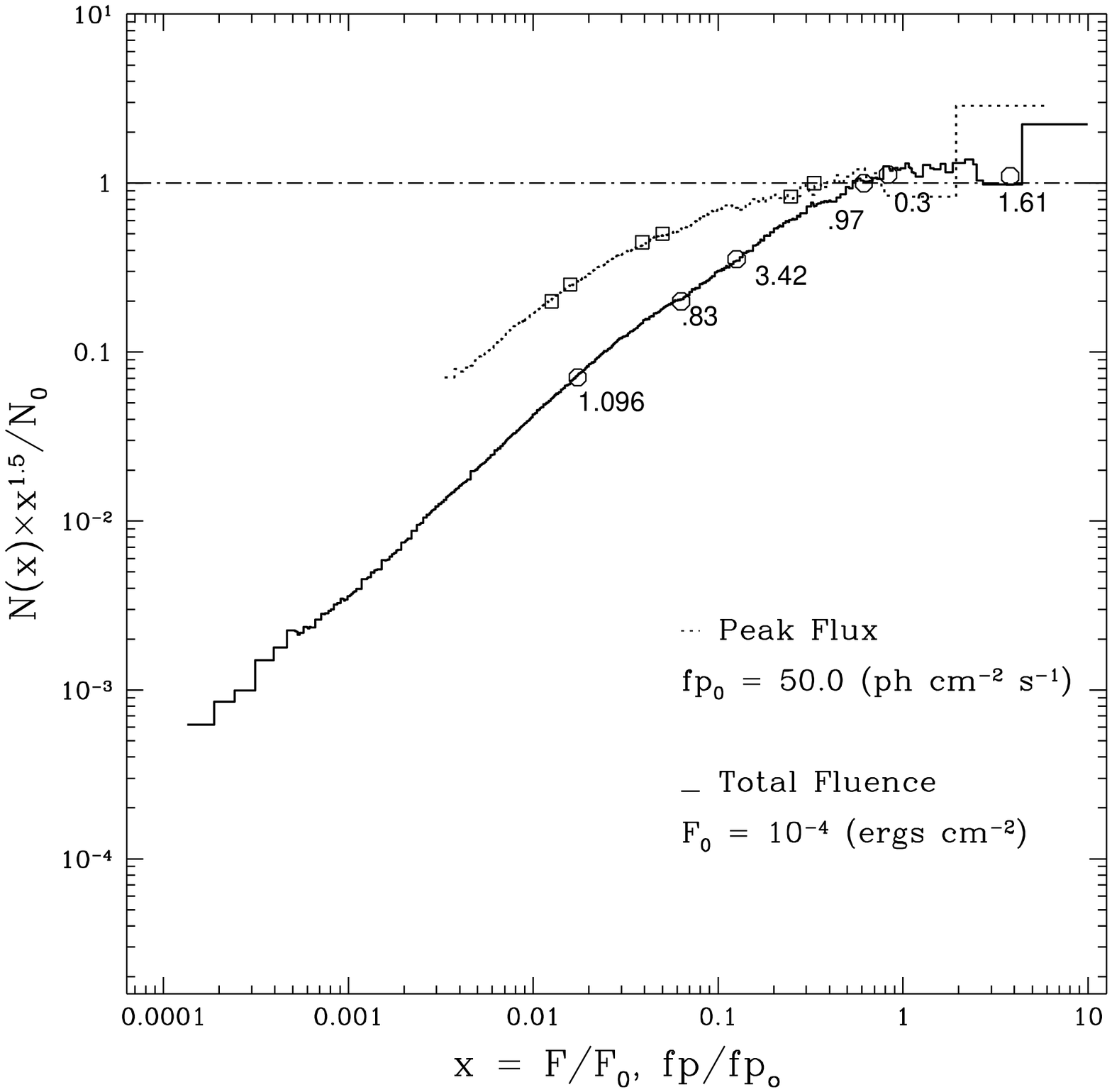,width=0.5\textwidth,height=0.5\textwidth}
}
\caption{\, {\bf  Left Panel:\,} Comparison of the distributions of $E_p$ 
observed by GINGA (Strohmeyer et al. 1997), SMM (Harris \& Share 1998) and 
BATSE (Mallozzi et al. 1998, the current sample).
{\bf Right Panel:\, }  Distribution of the total fluence and peak flux, with 
the Euclidean part taken out, with the known redshifts appropriately 
positioned.
}
\label{dist}
\end{figure}

Once the correlation between burst parameters is known, we can remove it and
use the techniques described in \S 2 to get an accurate estimation of the true
distributions of each parameter.  In particular, we can use these methods to
explore the truncation of the $E_{p}$ distribution.  However, beginning with a
narrow $E_{p}$ distribution for only the brightest bursts ($C_{p} \gg C_{lim}$),
we find that truncation plays a small role.  Not only do we need a more
complete sample of spectral fits to really explore this problem, but we need
to account for subtle effects of each burst's detector response matrix when
estimating the truncation. However, as shown in LP99 a more complete and well 
defined sample shows that the qualitative effects of the truncation is to 
produce an observed distribution that is significantly narrower than the 
actual distributuions. This behavior has also been seen in the simulations by 
Brainerd et al. (1999), where a broad (power law) parent distribution gives 
rise to a Gaussian like observed distribution and a narrow parent distribution 
leads to a somewhat narrower observed distribution. 

We can also use the available data from other instruments (sensitive to
different energy ranges than BATSE) to better understand the true $E_{p}$
distribution.  The left panel of Figure 8
superposes the BATSE, SMM,
and GINGA $E_{p}$ distributions.  The GINGA data (sensitive to lower energies
than BATSE) and SMM data (sensitive to higher energies than BATSE) show that
there are indeed a significant number of bursts outside the BATSE trigger
range as predicted by our methods.  Again, a more complete sample of spectral
fits along with the details of the bursts' DRMs are needed to really get a
handle on the  distribution of $E_{p}$.  However, the results from SMM and GINGA
certainly indicate that a raw correlation analysis without accounting for
truncation effects can lead to misleading results.

The right panel of Figure 8 shows the cumulative distribution (modulo 
the Euclidean part) of the total fluence and the peak flux with the redshifts 
of the known sources appropriately located. Significant deviations of these 
distributions from the Euclidean case are expected to begin at redshifts of 
less than one. The position of the redshifts along with the scatter diagram 
seen in Figure 1 once again point out the importance of the intrinsic 
dispersion vis-a-vis that expected from the cosmological effects.

\section{SUMMARY AND CONCLUSIONS}

Localization of GRBs by BeppoSAX, and the discovery
of optical afterglows and underlying galaxies with
measured redshifts has put the cosmological origin of GRBs
on firm footing.   The question which naturally arises is
to what degree the redshift distribution of the sources
affects the distributions and correlations between observed quantities,
and to what extent one can deduce the redshift distribution
from the latter.  There have been several suggestions of the existence
of cosmological redshift signatures in the temporal and spectral
characteristics of GRBs.  In this paper, we have expanded the search
for the cosmological signatures to other temporal and spectral features
in order to determine the validity of the claimed signatures, and the
extent of the influence of the redshift distribution on observables.
 
 We first point out that the observed redshifts when combined with
 the log$N$-log$S$ relation (Figs. 1 and 8) show the presence of a broad
 ``luminosity function''  not only at gamma-ray range (see, e.g. Stern et al. 1999)
 but at all photon energies. This should make
 the detection of cosmological signatures difficult, requiring 
 a careful analysis of observational and data analysis selection
 biases.  In \S 2, we discuss methods we have developed for determination
 of the selection biases and statistical procedures for accounting for
 the data truncations arising from these biases.
 
  We then present results from our study of temporal characteristics 
  of GRBs intended to investigate whether 
  the reported anticorrelation between
  the peak fluxes and some measure of duration of the bursts (e.g. 
  Norris et al 1994, Stern et al. 1997,
  Mitrafanov 1998, Deng \& Schaefer 1999) is 
  caused by cosmological ``time dilation''
  or is an intrinsic property of the bursts.  Using
  the TTS data and a different analysis than employed by the above authors,
   we find 
  the following results:
  
  $\bullet$  We confirm the above {\bf anticorrelation} between the peak
  flux and several measures of duration of bursts. However, we point out that
  the observed
  anticorrelation among bursts is too strong compared
  to what is expected cosmologically, particularly if most bursts
  have high redshifts.
  
  $\bullet$  Anticorrelations similiar to the above among bursts are seen
  among pulses in individual bursts.  We find a statistically significant
  excess of anticorrelations over correlations between pulse
  amplitude (peak pulse counts) and pulse
  duration.  This is especially significant for bursts when the correlation 
  coefficient between the above two quantities is strong.
  Unlike the anticorrelation found among bursts, the anticorrelation
  among pulses of individual bursts cannot be attributed to cosmological
  time dilation and must
  be intrinsic.  This raises the possibility that both effects
  are intrinsic.  
  
  $\bullet$ We find a strong {\bf correlation} between various meausres
  of fluence and duration among bursts, confirming the earlier results
  of Lee and Petrosian (1996, 1997).  We also find a strong {\bf correlation}
  between total counts or fluence and duration of pulses in individual
  bursts.  Such correlations are not expected
  from redshift effects and must be intrinsic.  The fact that 
  amplitude, duration and fluences of pulses have complex correlations
  is an indication that neither the peak luminosity nor the total
  energy of pulses are standard candles.  When this is extended to the 
  whole burst, it can explain the absence of a Hubble Law in Figure 1.
  
  Some or all of the above relations may be due to our methodology
  of defining pulses, and determining fitting parameters.  
  For example, one would expect a more likely loss of short weak
  bursts than strong ones.  To answer this question, we have simulated
  a representative sample of bursts and followed our pulse fitting  and
  analysis procedure used for BATSE bursts.  We do find some biases
  which can introduce anticorrelation between amplitude and duration when
  there is none, or weaken a strong correlation between fluence and
  duration of pulses.  However, these effects are weaker
  than what is actually observed, especially when we compare simulated and 
actual bursts
  for which the trends are strongest
  and statistically more reliable.
  
  Next we consider the claim of the presence of
  cosmological signatures, particularly the spectral redshift
  of the break energy, in the spectra of GRBs, most of which can
  be fitted to various forms approximating a broken power
  law (Mallozzi et al. 1996 and 1998, Mitrofanov et al. 1999).  This
  claim is based on a correlation between the break energy and peak
  flux, which is assumed to be a good measure of distance
  or redshift.  This requires a narrow distribution of peak
  luminosities of bursts, which - as discussed above - does not seem
  to be the case.  We believe that selection effects can play an 
  important role and possibly produce false correlations.  To clarify this 
  situation, 
  we have selected a subsample of spectral fits to bursts (from a larger
  sample kindly
  provided by Dr. Mallozzi) which is nearly complete within well
  defined thresholds.  Using the techniques described in \S 2, we have 
  found the following results:
  
  $\bullet$ We find a very strong correlation between several measures
  of burst fluence and spectral break energy (or $E_{p}$, the peak
  energy of the $\nu F_{\nu}$ spectrum), but only a weak correlation
  between peak flux and $E_{p}$.
  This apparent contradiction with the Mallozzi et al. result could arise
  from the fact that in obtaining a well defined sample, we are
  limited to the brightest burst, which - according to Mallozzi et al. (1996, 
1998)-
  show only a weak correlation.  That is, most of the claimed correlation
  between peak flux and $E_{p}$ comes from weaker bursts.  Alternatively,
  it could be due to a improper accounting of the selection effects present
  in the data and analysis.  
  
  $\bullet$  A correlation between fluence and $E_{p}$ is expected
  from redshifts effects.  We have quantitatively tested to see if
  the observed correlation could be attributed to cosmological effects.  Our
  tests show that this could be the case for
  standard candle radiated energy
  and for a narrow intrinsic distribution of $E_{p}$.  Neither one of 
  these requirements seem reasonable.  For more reasonable models, e.g.
  a power law distrbituion in the radiated energy with a broad
  range ($>$ two orders of magnitude as in Figure 1), the expected
  correlation from cosmological effects is weakened considerably.  Hence, the
  observed
  correlations must be primarily intrinsic to the radiation processes 
  at the source.
  
    In summary, we see very little direct signs of
    cosmological redshift effects in the temporal and spectral
    properties.  The bulk of the correlations we have described
    must be intrinsic to the source.  A corollary of this is that
    the claimed hardness-duration relation (Kouvelioutou et al. 1998)
    most likely is also intrinsic to the source.  All of these intrinsic
    correlations must be explained by the energy release
    and radiation processes at the burst.

\end{document}